\documentclass[aps,prl,twocolumn,letterpaper,superscriptaddress]{revtex4-1}

\usepackage[colorlinks=true,urlcolor=blue,citecolor=blue,linkcolor=blue]{hyperref}
\usepackage[all]{hypcap}

\usepackage{graphicx}
\usepackage{dcolumn}
\usepackage{amsmath}
\usepackage{amssymb}
\usepackage{subfigure}
\usepackage{float}
\usepackage{epstopdf}
\usepackage{natbib}
\usepackage{color}

\begin{document}

\newcommand{\dtau}{\partial_\tau}

\newcommand{\MPM}{MPM}
\newcommand{\MPMs}{MPMs}

\newcommand{\eqnref}[1]{(\ref{#1})}

\newcommand{\ket}[1]{\left| #1 \right\rangle}
\newcommand{\bra}[1]{\left\langle #1 \right|}

\title{Topologically protected braiding in a single wire using Floquet Majorana modes}

\author{Bela Bauer}
\affiliation{Station Q, Microsoft
  Corporation, Santa Barbara, California 93106 USA}
\author{T. Pereg-Barnea}
\affiliation{Department of Physics, McGill University, Montr{\'e}al, Qu{\'e}bec, Canada}
\affiliation{Department of Condensed Matter Physics, Weizmann Institute of Science, Rehovot 76100, Israel}
\author{Torsten Karzig}
\affiliation{Station Q, Microsoft
  Corporation, Santa Barbara, California 93106 USA}
\author{Maria-Theresa Rieder}
\affiliation{Department of Condensed Matter Physics, Weizmann Institute of Science, Rehovot 76100, Israel}
\author{Gil Refael}
\affiliation{Walter Burke Institute for Theoretical Physics and Institute for Quantum Information and Matter,
California Institute of Technology, Pasadena, California 91125 USA}
\affiliation{Department of Physics, California Institute of Technology, Pasadena, California 91125 USA}
\author{Erez Berg}
\affiliation{Department of Condensed Matter Physics, Weizmann Institute of Science, Rehovot 76100, Israel}
\affiliation{Department of Physics, James Franck Institute, University of Chicago, Chicago, Illinois 60637 USA}
\author{Yuval Oreg}
\affiliation{Department of Condensed Matter Physics, Weizmann Institute of Science, Rehovot 76100, Israel}

\date{\today}

\newcommand{\etalcomma}{\textit{et al.,\ }}

\definecolor{rewrite}{RGB}{228,0,124}
\definecolor{note}{RGB}{0,153,76}
\definecolor{cite}{RGB}{153,31,255}
\definecolor{check}{RGB}{228,0,0}
\definecolor{cut}{RGB}{255,153,51}
\begin{abstract}
Majorana zero modes are a promising platform for topologically protected quantum information processing. Their non-Abelian nature, which is key for performing quantum gates, is most prominently exhibited through braiding. While originally formulated for two-dimensional (2d) systems, it has been shown that braiding can also be realized using one-dimensional (1d) wires by forming an essentially two-dimensional network. Here, we show that in driven systems far from equilibrium, one can do away with the second spatial dimension altogether by instead using quasienergy as the second dimension. To realize this, we use a Floquet topological superconductor which can exhibit Majorana modes at two special eigenvalues of the evolution operator, $0$ and $\pi$, and thus can realize four Majorana modes in a single, driven quantum wire. We describe and numerically evaluate a protocol that realizes a topologically protected exchange of two Majorana zero modes in a single wire by adiabatically modulating the Floquet drive and using the $\pi$ modes as auxiliary degrees of freedom.
\end{abstract}
\maketitle

Non-equilibrium systems have recently been shown to host a variety of novel phenomena with no equilibrium system equivalent. One of the early examples was discussed in Ref.~\onlinecite{Jiang}, which demonstrated that a driven $p$-wave superconducting wire can possess not only the well-known Majorana zero modes (MZMs) at zero energy~\cite{Kitaev2001,Motrunich2001}, but also so-called Majorana $\pi$ modes (\MPMs) at frequency $\omega/2$, with $\omega$ the frequency of the external drive.
These are but an example of a broader class of anomalous Floquet topological phases~\cite{Kitagawa2010,Rudner2013}, with no analogue in static (time-independent) systems. Other examples include Floquet symmetry-protected topological (Floquet-SPT) phases~\cite{vonKeyserlingk16a,Else16a,Potter16,Roy16}, and the closely related time-crystals~\cite{Wilczek12,Shapere12,Khemani2016,Else2016,Potter2016,VonKeyserlingk2016a}, where periodically driven interacting and disordered systems show a response at a multiple of the drive period. In all these systems, discrete time-translation symmetry
protects novel quantum states.

It is natural to ask whether the topological degrees of freedom that emerge in driven systems can be used to supplement equilibrium topological phases. Particularly interesting are Majorana zero modes ~\cite{Alicea2012,Beenakker2013,DasSarma2015,Lutchyn17,Aguado17}. It is well-known that they exhibit non-Abelian statistics: When several MZMs are present, the many-body ground state becomes degenerate, and adiabatically exchanging two well-separated MZMs carries out a non-trivial unitary transformation within the ground state manifold~\cite{Ivanov,Mariani}. Such braiding operations form the basis of topological quantum computation~\cite{Kitaev,Nayak}. Physically, MZMs are realized as zero-energy excitations in one-~\cite{Kitaev2001, Lutchyn,ORvO,Kouwenhoven,Yazdani} and two-~\cite{Sato,Lee,FuKane,DasSarma} dimensional topological superconductors. While these systems are of great interest for quantum computing, non-Abelian braiding itself remains a tantalizing fundamental effect, and demonstrating it would be a tremendous breakthrough. 

In the following, we show that \MPMs{} emerging in driven systems allow for remarkable new braiding protocols, going beyond what is possible in equilibrium systems. Strictly speaking, braiding is only possible beyond one spatial dimension: two quasi-particles cannot be exchanged on a single wire while being distant from each other. In this work, however, we show that in periodically driven systems, \emph{quasienergy} provides an additional synthetic dimension that can be used in conjunction with real space. Roughly speaking, the two kinds of Majorana states in Floquet superconductors -- MZMs and \MPMs{} -- live a parallel existence at two different frequencies. As pointed out first in Ref.~\cite{Jiang}, they are precisely decoupled from each other as long as the drive is invariant under time-translation by one period.  It follows that half-frequency pulses can be used to couple the MZMs and \MPMs{}~\footnote{This was previously discussed in the dual language of destroying the emerging $\mathbb{Z}_2$ symmetry in an Ising time crystal in, e.g., Ref.~\cite{Else2017}.}.

Refs.~\cite{Bomantara18,Bomantara18-1} also propose using a combination of MZMs and \MPMs{} as well as half-frequency pulses to simulate braiding operations. However, the scheme we perform here is a \emph{non-local} braid rather than a local operation at one end of the system. The non-locality of our scheme leads to topological protection against local perturbations.

{\it Floquet braiding}---We begin with a 1d topological superconductor, which
under a period-$T$ drive may enter a Floquet topological superconducting phase~\cite{Jiang,Liu2013}.
As a function of material and drive parameters, each edge of the system may have no MZMs, one MZM and/or another Majorana mode with energy at the Floquet zone boundary.  We denote this quasienergy by $\pi/T$ and refer to the corresponding Majorana mode as a Majorana $\pi$ mode (\MPM). A time-periodic system only allows quasienergies inside the Floquet zone, $-\pi/T\le\epsilon<\pi/T$. Therefore, particle-hole symmetry requires that Majorana modes come in pairs at all energies except zero and $\pi/T$, which is where unpaired Majorana modes can be found.  Moreover, as long as time periodicity is conserved, the MZMs and \MPMs{} do not hybridize even if their wavefunctions overlap in space. This property allows us to move them past each other and enables the procedure,
which does not
require any fine-tuning of the Hamiltonian or its time dependence~\footnote{For a protocol that relies on fine-tuning of the Hamiltonian, see Ref.~\cite{Chiu2015}.}.

There are several experimental schemes for MZM exchange. The simplest one is to physically move the MZMs~\cite{Alicea2}. Alternatively, consider a system made up of four MZMs
at fixed locations, but with tunable interactions between them~\cite{vanHeck,Hyart2013,Aasen2016,Karzig}; in this case, at any time
during the braid two of the four Majorana modes are strongly coupled, but the dominant coupling is changed
in a particular order to effectively perform a braid operation. Similarly, a sequence of 2-MZM measurements can be used to implement measurement-only variants of braiding \cite{Bonderson2008,Plugge2017,Karzig2017}. In either case, at least two quantum wires are required.

Our proposed braiding protocol is most closely akin to an approach
with four MZMs, of which two are coupled at any time. Our four states, however, are a pair of MZMs and another pair of \MPMs{}.
To introduce interactions between MZMs and \MPMs{}, we apply a time-dependent perturbation in restricted regions, thus locally breaking the time-translation 
symmetry that protects the \MPMs{}. We numerically confirm below that such a perturbation acts only locally even
though time-translation symmetry is a global symmetry. We then combine this with moving the
MZMs and \MPMs{} to achieve braiding.

\begin{figure}
\begin{minipage}[t]{.5\columnwidth}
	\vspace{0pt}
	\centering
	\includegraphics{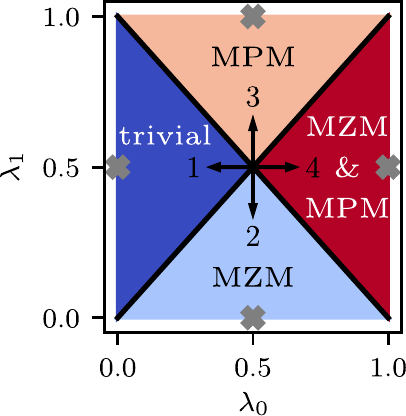}
\end{minipage}%
\begin{minipage}[t]{.5\columnwidth}
	\vspace{0.22in}
	\centering
	{\renewcommand{\arraystretch}{1.2}
    \begin{tabular}{|l||l|l|}
    \hline
    & $\lambda_0$ & $\lambda_1$\\
    \hline \hline
    1. & $2\lambda_0^1 = 1-\delta$ & $2\lambda_1^1 = 1$ \\
    \hline 
    2. & $2\lambda_0^2 = 1$ & $2\lambda_1^2 = 1-\delta$ \\
    \hline 
    3. & $2\lambda_0^3 = 1$ & $2\lambda_1^3 = 1+\delta$ \\
    \hline 
    4. & $2\lambda_0^4 = 1+\delta$ & $2\lambda_1^4 = 1$ \\
    \hline 
    \end{tabular}}
\end{minipage}
  \caption{
  \textit{Left:} Phase diagram of the Floquet system in terms of the strength of the topological (trivial) Hamiltonian $H_0 (H_1)$ in the two-part drive, see Eqs.~\eqref{eq:drive0},\eqref{eq:drive1}. It is possible to realize 4 phases characterized by the presence of zero or $\pi$ modes: 1) trivial 2) MZMs 3) \MPMs{} 4)  MZMs and \MPMs{}. The gray crosses mark the sweet spots of the corresponding phases with vanishing correlation lengths. \textit{Right:} Parameterization used in Eq.~\eqref{eqn:UF} to obtain the phase diagram. The parameter $\delta$ quantifies the distance to the critical point and the direction of increasing $\delta$ is indicated in the left panel.\label{fig:phasediagram}}
\end{figure}

{\it Two-part drive model}---Let us consider the Kitaev Hamiltonian:
\begin{multline}
H(\mu_i,w_i,\Delta_i) = \sum_i\left[-\mu_i c_i^\dagger c_i - \frac{w_i}{2} \left( c_i^\dagger c_{i+1} + \mathrm{h.c.} \right)\right] \\ + \sum_i \frac{\Delta_i}{2}  \left( c_i c_{i+1} + \mathrm{h.c.} \right)
\end{multline}
and construct the Floquet operator with period $T$
\begin{align} \label{eqn:UF}
U_F &= e^{-i H_0 T/2} e^{-i H_1 T/2} \\
H_0 &= H(\mu_i = 0, w_i = 2\pi \lambda_0/T , \Delta_i = -2\pi \lambda_0/T) \label{eq:drive0} \\
H_1 &= H(\mu_i = 2\pi \lambda_1/T, w_i = 0, \Delta_i = 0)\, \label{eq:drive1},
\end{align}
where $H_0$ is the Hamiltonian of a Kitaev chain at the ``sweet spot'' of the topological phase (see below) and $H_1$ is the Hamiltonian of a trivial phase with only chemical potential.
For couplings $\lambda_0,\lambda_1 \in [0,1]$ ($\hbar=1$ throughout), this gives rise to the phase diagram \cite{Khemani2016} (Fig.~\ref{fig:phasediagram}) with the
four phases characterized by the presence or absence of MZMs and \MPMs. Each phase contains a point of vanishing
correlation length, \textit{aka} `sweet spots,' where the MZM and/or MPM states are localized on a single site.  The sweet spots are indicated by the gray crosses in Fig.~\ref{fig:phasediagram}. These points are discussed in the Supplemental Material. A convenient choice of parameters is given by that of the right panel of Fig.~\ref{fig:phasediagram},
where the superscript refers to the phases as follows: 1--trivial phase, 2--MZM only, 3--\MPM{} only, and 4--both MZMs and \MPMs.
The parameter $\delta$ quantifies the distance of all the phases to the critical point and is connected to the correlation length, with $\delta=0$ corresponding to the critical point and $\delta=1$ to the points with vanishing correlation length (the black arrows in the left panel of Fig.~\ref{fig:phasediagram} indicate the direction of increasing $\delta$).
Throughout, we consider $U_F$ to encode an elementary Floquet cycle with period $T$. 

To implement the Floquet braiding protocol, consider an inhomogeneous systems, where different regions are in different phases with the possibility to move phase boundaries. Let $\vec{p}$
be a vector 
whose elements $p_i \in \lbrace 1,2,3,4 \rbrace$ indicate that the parameters of the bond $i$ correspond to phase $p_i$. We can then
generalize the Floquet drive of Eq.~\eqnref{eqn:UF} to the inhomogeneous case:
\begin{align} \label{eqn:UFinh}
U_F(\vec{p}) &= e^{-i H_0(\vec{p}) T/2} e^{-i H_1(\vec{p})T/2} \\
H_0(\vec{p}) &= H(\mu_i = 0, w_i = 2\pi \lambda_0^{p_i}/T, \Delta_i = -2\pi \lambda_0^{p_i}/T) \\
H_1(\vec{p}) &= H(\mu_i = 2\pi \lambda_1^{p_i}/T, w_i = 0, \Delta_i = 0).
\end{align}
In an inhomogeneous system, MZMs and \MPMs{} also form at the interfaces between phases with different
topological order. For example,
half of the system could be in phase 2 (MZM), and the other half in phase 4 (MZM and \MPM). In such a case, the MZMs will form at the end of the system, one \MPM{} will
form at one end of the system, and the other one in the middle of the system.

To move the spatial phase boundaries as a function of time, we interpolate between two different systems described by vectors $\vec{p}$ and $\vec{q}$ by continuously tuning a parameter $s \in [0,1]$ and applying
Floquet drives analogous to Eq.~\eqnref{eqn:UFinh}, but with $H_0 = (1-f(s)) H_0(\vec{p}) + f(s) H_0(\vec{q})$,
and similarly for $H_1$. Here, $f(s)$ is a function with $f(0)=0$ and $f(1)=1$; in our simulations, we choose
$f(s) = \sin(s \pi/2)^2$. We evolve from $s=0$ to $s=1$ over $N_s$ time steps.
For sufficiently large $N_s$, if the initial state of this operation is an eigenstate of $U_F(\vec{p})$, the final state will be an eigenstate of $U_F(\vec{q})$. This can be considered a version of adiabaticity for driven  systems~\cite{Breuer1989,Kitagawa2011,Weinberg17} and be understood by the formal relation between each $U_F$ to a Floquet Hamiltonian $H_F = i (\log U_F)/T$. The spectrum of $H_F$ corresponds to the quasi-energy spectrum of the Floquet unitary. We can therefore relate
the deformation from $U_F(\vec{p})$ to $U_F(\vec{q})$ to a deformation of the corresponding
Floquet Hamiltonian from $H_F(\vec{p})$ to $H_F(\vec{q})$. The adiabatic condition can then be formulated with respect to the quasienergy spectrum of $H_F$.
Dynamically changing the Floquet operator weakly breaks the time-translation symmetry that protects the \MPMs{} similar to how energy conservation is broken in time-dependent equilibrium systems. 
To reduce the corresponding errors in the braiding protocol, we choose a smooth evolution which strongly suppresses the $\pi/T$ components as $N_s$ becomes large except for the desired local perturbations discussed below.

{\it Local time-translation symmetry breaking}---As a final ingredient to our protocol, we need to be able to couple nearby MZMs and \MPMs{}. To explicitly
introduce such a coupling, we insert an operator $U_{\rm pert}$ after every two elementary Floquet cycles, thus changing $U_F^n$, to $\left( U_F^2 U_{\rm pert} \right)^{n/2}$. The coupling can be understood by considering that eigenvectors corresponding to quasi energies 0 and $\pi/T$ in $U_F$ all correspond to quasienergy 0 in $U_F^2$, and are therefore susceptible to perturbations. 
Importantly, if $U_{\rm pert}$ acts only in a specific
region of the system, it will only couple a pair of nearby MZMs and \MPMs{} in that region while leaving the ones far away unperturbed.

\begin{figure}
	\includegraphics{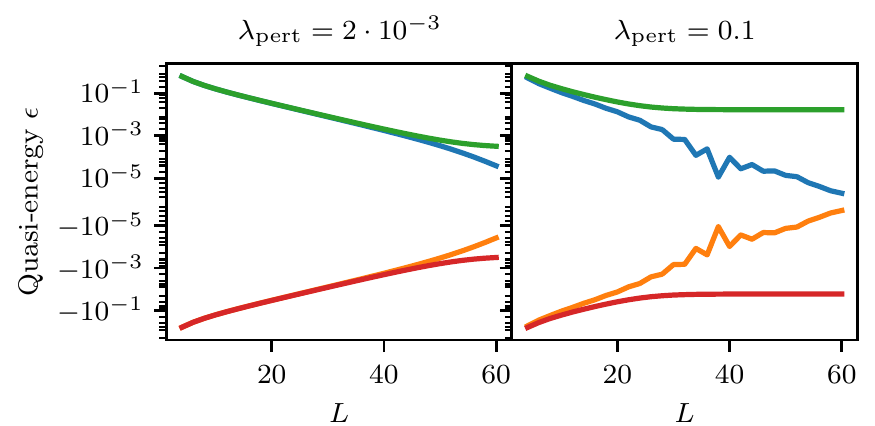}
	\caption{
		Quasi energies closest to zero for Floquet evolutions over two cycles for a system of length $L$ in phase 4 ($\delta = 0.09$) for different strengths of time-translational symmetry breaking perturbations applied the right end of the system. Left, right panel show the case of a very weak ($\lambda_{\rm pert}=2 \cdot 10^{-3}$)
		and moderate perturbation ($\lambda_{\rm pert}=0.1$), respectively. In the unperturbed 
		case, each level is two-fold degenerate corresponding
		to two MZMs and two \MPMs. In the perturbed case, since the pair of MZM and \MPM{} at the right end of the
		system is split, only the MZM and \MPM{} at the left end remains. Notice that when $\lambda_{\mathrm{pert}}\ne 0$ the period is doubled, and as a result the MZMs and MPMs both get folded to the vicinity of $\epsilon=0$.\label{fig:lttsb} }
\end{figure}

To confirm this picture, we turn to numerical simulations, which we perform using established techniques~\footnote{For an overview, see the Supplemental Material as well as Refs.~\cite{Wimmer2012,Bravyi2017,Bauer18}}.
We compute the spectrum of the operator $U_F(\delta)^2 U_{\rm pert}$, where $U_F(\delta)$ 
is the Floquet operator of Eq.~\eqnref{eqn:UF} with the parameters chosen inside phase 4 which exhibits both zero and $\pi$ modes, and $U_{\rm pert}$ acting only on one half of the system. Specifically, we choose
\begin{equation}
U_{\rm pert}(\lambda_{\rm pert}) = e^{i T H(T\mu_i = \lambda_{\rm pert},T w_i = \lambda_{\rm pert}, T\Delta_i = -\lambda_{\rm pert})},
\end{equation}
where $\mu_i$, $w_i$ and $\Delta_i$ are non-vanishing only in the right half of the system.
For $\lambda_{\rm pert} = 0$, time-translation symmetry for a single Floquet cycle is restored and the system will exhibit two localized and uncoupled modes at each end. However, when $\lambda_{\rm pert} > 0$, the (0 and $\pi$)  modes at the right end split, while the MZM and \MPM{} at the left remain as the only unsplit modes.
This behavior is reflected in the spectrum shown in Fig.~\ref{fig:lttsb}, which shows the lowest (in absolute value)
quasi-energies of $U_F(\delta)^2 U_{\rm pert}$ for two choices of $\lambda_{\rm pert}$.
Due to particle-hole symmetry, the positive and negative quasi energies mirror each other.
For the unperturbed case, $\lambda_{\rm pert} = 0$, we find that four eigenvalues approach
zero exponentially as the system size is increased. Upon perturbing the system, two of them saturate to a value of order
$\lambda_{\rm pert}$, while the others continues to decrease exponentially with the same exponent that governed the unperturbed case.

\begin{figure}
\begin{minipage}[t]{.6\columnwidth}
\vspace{-35mm}
\centering
\includegraphics[trim=0.5cm 0.5cm 0.5cm 0.65cm]{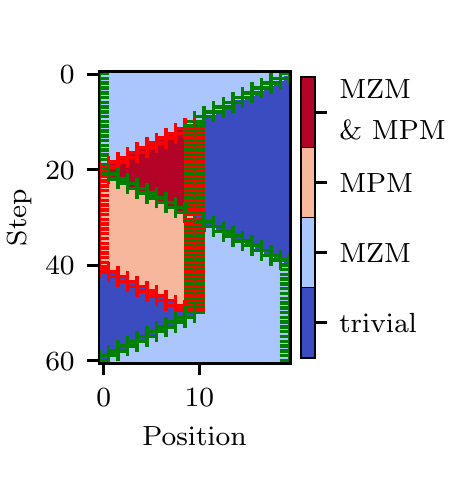}
\end{minipage}%
\begin{minipage}[t]{0.4\columnwidth}
\centering
\includegraphics[width=0.5\columnwidth]{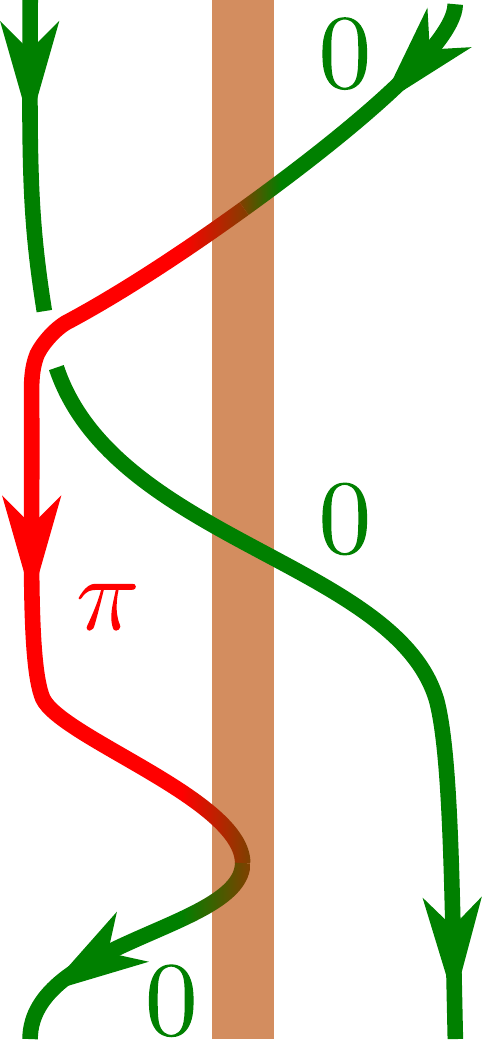}
\end{minipage} 
\caption{\textit{Left:} Full braid protocol for a system of $L=20$ sites. The colors correspond to different phases; green (red) crosses indicate the locations of MZMs (\MPMs{}). \textit{Right:} Schematic representation of the braiding process of two MZMs. In the center region it is possible to convert between MZMs (denoted by 0) \MPMs{} (denoted by $\pi$). After the right MZM has been converted into a \MPM{} is can be safely moved past the left MZM in the region where time-translational symmetry is preserved.    \label{fig:protocol} }
\end{figure}

{\it Braiding protocol}---We now turn to the full braid protocol. We start and end in a configuration where the entire system is in the regular, undriven,
Kitaev phase, exhibiting MZMs at the system's edge. This allows state preparation in an undriven system. We then turn on the Floquet drive to
perform a braid operation by following the steps in Fig.~\ref{fig:protocol}. Since all the Floquet-drive phases ~\eqnref{eqn:UF} are gapped around the respective 0 or $\pi$ modes, and the protocol never drives extended regions of the system through the phase transition at once, the Floquet quasienergy spectrum at each step of the evolution remains gapped. Therefore adiabaticity is maintained even in the thermodynamic limit by choosing $N_s$ which interpolates the move of the phase boundary by one site sufficiently large.

Throughout the evolution, the system contains at least a pair of MZMs, and, at intermediate stages, an additional a pair of \MPMs{}. In the case where both MZMs and \MPMs{} and hence a total of four modes are present, we need to fix which pair  encodes the quantum information.
To achieve this, we apply a local time-translation-symmetry-breaking perturbation in a region in the middle of the system. Therefore, when both an MZM and an \MPM{} are in the middle, they are split to finite energy and only two low-energy modes remain, which thus carry the encoded quantum state. When three modes, e.g. two \MPMs{} and an MZM, are in the perturbed regime, one mode (which is a linear combination of the three modes) remains unperturbed while two are split away to finite energy. This enables us to effectively convert a MZM to a \MPM{} mode and vice versa as indicated in Fig.~\ref{fig:protocol}.

\begin{figure}
  \includegraphics{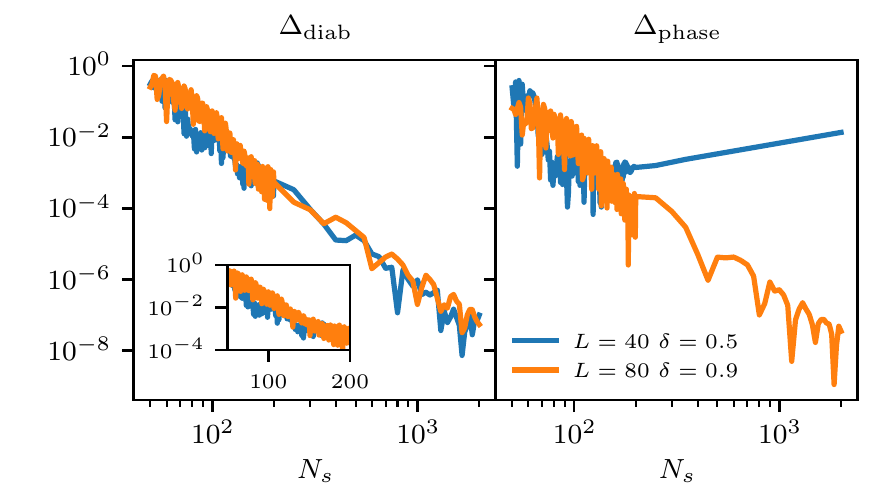}
  \caption{Errors in the braid protocol, measured by the deviation from unitarity of the evolution in
  the low-energy subspace (left panel) and deviation in the applied phase (right panel, see main text
  for definitions of $\Delta_{\rm diab}$ and $\Delta_{\rm phase}$) as a function of the number of 
  interpolation steps between stages of the protocol. In the limit of $N_s \rightarrow \infty$,
  adiabaticity is recovered. The errors generally vanish with a lower-law, however for fast protocols
  ($N_s < 200$) an exponential transient behavior is observed. In the phase error, the dependence on
  $N_s$ is non-monotonic: for sufficiently slow protocols, the evolution becomes adiabatic with respect
  to the residual finite-size splitting of Majorana modes.
  Parameters used are $\lambda_{\rm pert}=0.2$.
  \label{fig:diab} }
\end{figure}

A subtle point arises if both a perturbation that breaks time-translation symmetry is present, and the Floquet drive is slowly changed to move phase boundaries as described above. In that case the Floquet unitary of a single cycle is $U_F(s)$ with $s$ a slowly changing parameter such that consecutive cycles are described by $U_F(s') U_F(s)$ with $s'-s=1/N_s$. When adding a perturbation $U_{\rm pert}$ to this, it is important that the parameter $s$ is still changed in every step, i.e. the perturbed evolution over two cycles is $U_{\rm pert} U_F(s') U_F(s)$. The perturbation will still be effective as long as $U_F(s)$ and $U_F(s')$ are sufficiently close.
While it may appear more natural to change the parameter $s$ only every two cycles when inserting the perturbation, this inadvertently induces an additional half-frequency perturbation. While the strength of this accidental perturbation vanishes in the adiabatic limit $N_s\rightarrow \infty$ where $s$ is changed only infinitesimally, it is also applied a diverging number of times in that limit, and thus a net effect remains. The protocol then exhibits non-universal corrections even when performed in the adiabatic limit. Similar corrections may be explicitly exploited to perform certain geometric quantum gates \cite{Bomantara18,Bomantara18-1}.

{\it Numerical results}---A numerical implementation of the dynamical braiding is summarized in Fig.~\ref{fig:diab}. Since the Hamiltonian is quadratic, the evolution of operators of the form $\vec{v}\cdot \vec{\gamma}$, where $\vec{\gamma}$ is a vector of Majorana operators such that $2c_i = \gamma_{2i-1} + i \gamma_{2i}$, can be represented by an orthogonal matrix ${\bar{U}}$. Over the entire process, $\vec{v}\cdot \vec{\gamma}$ evolves into $({\bar{U}} \vec{v})\cdot \vec{\gamma}$ (see Supplementary Information for details). 

To define the
relevant error measures, let 
$\gamma_{1,2} = \vec{v}_{1,2} \cdot \vec{\gamma}$ be initial (and final) MZMs. Then, we compute the $2 \times 2$ matrix $(U_r)_{\alpha,\beta} = \vec{v}^{T}_\alpha \bar{U} \vec{v}^{\vphantom{T}}_\beta$ ($\alpha,\beta=1,2$),
which encapsulates how the entire time evolution acts on the low-energy Majorana subspace. In the ideal
limit, $U_r = i \sigma^y$, where $\sigma^y$ denotes the usual Pauli matrix. We quantify deviations from this using two measures:
$\Delta_{\rm diab} = | U_r^\dagger U_r - 1 |$ captures deviations from unitarity, in particular diabatic corrections
that excite fermions from the low-energy subspace to the excited states.
Secondly, we compute the two eigenvalues of $U_r$ as $(r_1 e^{i \phi_1}, r_2 e^{i \phi_2})$. In the ideal
case, we expect $r_1=r_2=1$ and $\phi_1 = -\pi/2$, $\phi_2 = \pi/2$. We define deviations from this as
$\Delta_{\rm phase} = |\phi_1+\pi/2|+|\phi_2-\pi/2|$, where we sort eigenvalues such that
$\phi_1 \geq \phi_2$. Both measures are chosen to be independent of the
basis choice for the Majorana subpsace since it is not unique in the case when they are exactly degenerate.

Fig.~\ref{fig:diab} shows that increasing $N_s$ to perform a slower protocol
improves the errors. At short times, the accuracy improves exponentially,
while at long times a power-law behavior is observed, consistent with the 
non-analytic time-dependence of the driving Hamiltonian.
Interestingly, the two error measures can exhibit qualitatively different behavior, as shown
in the long-time behavior for $L=40$, $\delta=0.5$: while the diabatic corrections continue
to decrease, the error in the applied phase reaches a minimum value beyond which it increases again.
This occurs because very slow
protocols resolve the splitting of the low-energy manifold. For larger system sizes, such as $L=80$ and
$\delta=0.9$, this crossover would occur at much slower protocol times (larger $N_s$). In most relevant parameter regimes, the error is dominated by diabatic corrections
and not finite-size corrections, i.e. the error is independent of system size for all but the
smallest systems. Details of the dependence of $\Delta_{\rm diab}$ on other parameters such as $\delta$ and $\lambda_{\rm pert}$ can be found in the Supplemental Material.

{\it Topological protection \& Outlook}---To conclude, we discuss in what sense braiding as described here is topologically protected. 
Just as many other new phenomena in periodically driven systems, \MPMs{} are protected by time-translation symmetry. Therefore, braiding of \MPMs{} is topologically protected only if no processes that break the periodicity of the drive are present. A subtle issue is that the braid process itself breaks time-translation symmetry and thus gives rise to dynamical corrections, but as we have shown above these can be systematically suppressed by adiabatically changing the drive parameters. Similar diabatic errors may also occur in the braiding of MZMs if operations are performed away from the adiabatic limit~\cite{Karzig2013,Scheurer2013,Amorim2014,Karzig2015,Karzig2015b,Pedrocchi2015,Pedrocchi2015b,Knapp2016,Hell2016,Sekania2017,Rahmani2017,Bauer18}. 

Importantly, unlike other symmetries that can give rise to multiple MZMs in a single wire, our Floquet approach does not require careful tuning of the instantaneous Hamiltonian. Thus it is much more experimentally accessible. We provide a perspective towards such realizations in systems based on superconducting quantum dot chains \cite{Sau12,Fulga13,Su17} in the Supplemental Material, where in particular we discuss a model that is able to implement the same behavior but requires time-dependent control of only a single parameter. Perhaps the simplest realization, however, would be using a quantum wire proximity coupled to two superconductors, one grounded, and the other at a finite voltage. The AC Josephson effect gives rise to the time dependence leading to MPM's \cite{Peng2018}. 

An important caveat is that we relied on the absence of heating. While this assumption is appropriate for the non-interacting limit, it is well-known that driven interacting systems generically heat to infinite temperature~\cite{DAlessio2014,Lazarides2014,Ponte15a}. However, there are known mechanisms such as many-body localization~\cite{Abanin2014,Ponte15a,Ponte15b,Lazarides15}
as well as the pre-thermalization ~\cite{Abanin2015,Abanin2015a,Abanin2015b,Kuwahara2015,Mori2016,bukov2015,canovi2016,bukov2016} which can be used to avoid heating and stabilize the results discussed here. The details of this interacting scenario are an open question left to future work.

\acknowledgements

This work was supported by NSERC DG (TPB), the BSF and ISF grants
and by the European Research Council under the European Community’s Seventh Framework
Program (FP7/2007–2013)/ERC - Grant agreement MUNATOP-340210.
YO and EB acknowledge support from CRC 183 of the Deutsche Forschungsgemeinschaft.
We are also grateful for the hospitality of the Aspen Center for Physics, which is supported by National Science Foundation grant PHY-1607761, and where part of the work was done. 
\bibliography{Exchange}

\clearpage








\newpage

\section{Supplemental Material}

\makeatletter
\newcommand{\xRightarrow}[2][]{\ext@arrow 0359\Rightarrowfill@{#1}{#2}}
\makeatother

\subsection{The sweet spots}
\begin{figure}[h]
	\includegraphics[width=0.9\columnwidth]{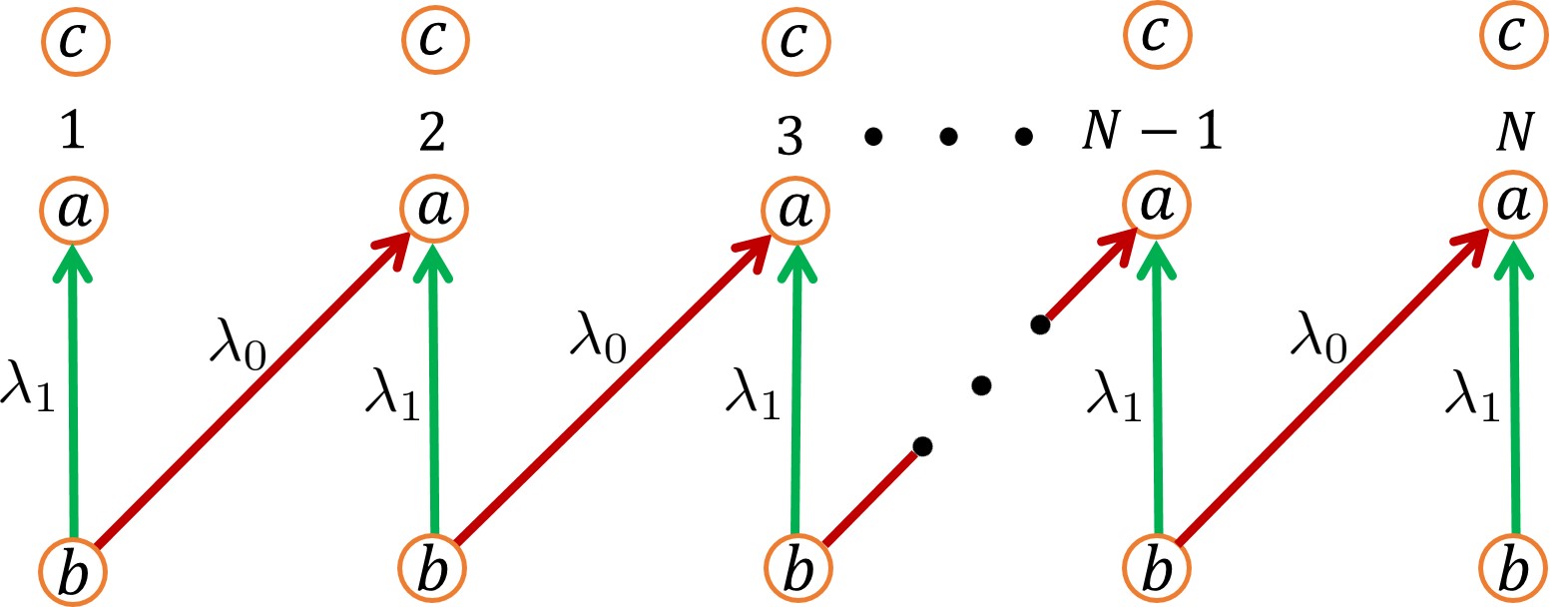}
	\caption{The Hamiltonian Eq. (\ref{abh}). Red links represent $H_0$, and green links  $H_1$.  An application of $U_0 \equiv e^{-i H_0 T/2}$ with $\lambda_0=1/2$ exchanges the positions of the MZM $b_n$ and $a_{n+1}$ yielding $b_{n} \rightarrow a_{n+1}$ and $a_{n+1} \rightarrow -b_{n}$. Similarly, $U_1 \equiv e^{-i H_1 T/2}$ with $\lambda_1=1/2$ carries out the transformation $a_n \rightarrow b_{n}$ and $b_{n} \rightarrow -a_n$. The four sweet spots (indicated by crosses in the left panel of  Fig.~\ref{fig:phasediagram} of the main text) are obtained by successive application of $U_0$, $U_1$ and the identity operator (i.e., a vanishing Hamiltonian for time $T/2$) $I=e^{-i H_0 T/2}=e^{-i H_1 T/2}$ with $\lambda_0=\lambda_1=0$. The trivial phase is obtained with the application of $I U_1$. The MZM phase, which contain MZM only, is obtained with the application of $U_0I$. The MPM phase, which has only $\pi$ modes, is obtained with the application of $U_0 U_1^2$. Finally, the MZM and MPM phase, having both zero and $\pi$ Majorana modes is obtained with the application of $U_0^2 U_1$. Notice that the application of $U_{0(1)}$ twice is equivalent to taking $e^{- i H_{0(1)} T/2}$ but now with $\lambda_{0(1)}=1$.  \label{fig:MajoChain}}
\end{figure}

In this section we revisit the `sweet spots' mentioned in the main text. The `sweet spots' describe locations in the phase diagram where both MZM and MPM are localized on one or two sites, i.e. the correlation length vanishes. Let us provide a simple analytical approach to deriving these sweet spots. First, let us denote each Dirac Fermion operator $c$ by two Majorana operators $a$ and $b$ on each site. Formally we substitute $2c_n=a_n+i b_n$ with $\{a_n,b_m\}=0$ for all $n$ and $m$ and $a_n^2=b_n^2=1$. (These are related to $\vec{\gamma}$ introduced in the main text by $a_n = \gamma_{2n-1}$, $b_n = \gamma_{2n}$.)
Then the model becomes
\begin{equation}
H_0=-i \frac{\pi}{T}\lambda_0 \sum_{n=1}^{N-1} a_{n+1} b_{n} \;\; H_1=-i \frac{\pi}{T}\lambda_1\sum_{n=1}^N a_n b_n,.\label{abh}
\end{equation}
This model is depicted in Fig.~\ref{fig:MajoChain}.
 
We note that application of $e^{-i H_0 T/2}$ with $\lambda_0=1/2$ exchanges the positions of $b_n$ and $a_{n+1}$ for $n=1,2,\dots, N-1$.
 Indeed, defining
 \begin{equation}
 U_0 \equiv e^{\frac{\pi}{4} \sum_{n=1}^{N-1}  b_n a_{n+1}} =\Pi_{n=1}^{N-1} B_0^n,\; B_0^n=e^{\frac{\pi}{4} b_n a_{n+1}},
  \end{equation}
  one can readily check that
  \begin{equation}
  {B_0^n}^\dagger b_n B_0^n= a_{n+1}\;\text{ and}\;  {B^n_0}^\dagger a_{n+1} B_0^n= -b_n.
   \end{equation}
   Notice that $B_0^n$ and $B_0^m$ commute for $n \ne m$. Similarly
   \begin{equation}
  U_1 \equiv e^{\frac{\pi}{4} \sum_{n=1}^{N} a_n b_{n}} =\Pi_{n=1}^{N} B_1^n,\;\text{ and}\; B_1^n=e^{\frac{\pi}{4}a_n b_n},
   \end{equation}
  and
  \begin{equation}
 {B_1^n}^\dagger b_n B_1^n= a_n,\; \text{and}\;  {B^n_1}^\dagger a_n B^n_1= -b_n.
   \end{equation}
   These operations are depicted in Fig.~\ref{fig:MajoChain}. The arrow indicates which Majorana operator acquires the minus sign. For example, in the application of $U_1$ (green arrows in Fig.~\ref{fig:MajoChain}), $b_n \rightarrow a_n$ as the arrow directed from $b_n$ to $a_n$ while $a_n \rightarrow -b_n$.

 Using these observations, and the identity operator (a vanishing Hamiltonian for $T/2$) $I=e^{-i H_0 T/2}=e^{-i H_1 T/2}$ with $\lambda_0=\lambda_1=0$, it is now straightforward to identified the operation of the Floquet operator $U_F =e^{-i H_0 T/2} e^{-i H_1 T/2}$ at the sweet spots in the various phases. 
\subsubsection{1. Trivial}
The trivial phase is obtained with the application of $I U_1$ (corresponding to $\lambda_0=0, \lambda_1=1/2$) then:
 $$a_n \xrightarrow{I}{} a_n \xrightarrow{U_1} -b_n \text{ and } b_n \xrightarrow{I}{} b_n \xrightarrow{U_1} a_n $$
   for $n=1,\dots, N$.
  So that in the subspace spanned by $a_n$ and $b_n$ the operator $\vec{v}_n \cdot  (a_n, b_n)^T$ evolves into $(\bar{U}_F \vec{v}_n)\cdot (a_n,b_n)^T$, with
  $$\bar{U}_F=\left(
          \begin{array}{cc}
            0 & 1 \\
            -1 & 0 \\
          \end{array}
        \right)=i \sigma_y,$$
  having eigenvalues $\pm i= e^{\pm i \epsilon_n T}$ with quasi-energies $\epsilon_n=\pm \pi/(2 T)$ for all $n$, which are not corresponding to Majorana modes, occurring at quasi-energies zero or $\pi/T$. 
\subsubsection{2. MZM}
The phase with MZM only at the two ends of the wire is obtained with the application of $U_0 I$ (corresponding to $\lambda_0=1/2, \lambda_1=0$) then:
   $$a_{n+1} \xrightarrow{U_0}{} -b_{n} \xrightarrow{I} -b_{n} \text{ and } b_n \xrightarrow{U_0}{} a_{n+1} \xrightarrow{I} a_{n+1}, $$
  for $n=1,\dots N-1$.
   So that in the subspace spanned by $a_{n+1}$ and $b_n$ (for $n=1,\dots, N-1$) we find, similarly to the trivial case, quasi-energies $\pm \pi/(2T)$,
  but the Majorana operators $a_1$ and $b_N$ remain unchanged, establishing the presence of two MZM modes which are localized on one site.
  In the subspace spanned by $a_1$ and $b_N$ we find that $\bar{U}_F$ is the identity matrix
  with eigenvalues $1=e^{i \epsilon T}$, and two quasi-energies $\epsilon=0$.
\subsubsection{3. MPM}
The phase with MPM only at the two ends of the wire is obtained with the application of $U_0 U_1^2$ corresponding to $\lambda_0=1/2, \lambda_1=1$; notice that the application of $U_{0(1)}$ twice is equivalent to taking $e^{- i H_{0(1)} T/2}$ with $\lambda_{0(1)}=1$, and results in the multiplication of the Majorana operator by $-1$. Then,
       \begin{eqnarray*}
   a_{n+1} &\xrightarrow{U_0}{}& -b_{n} \xrightarrow{(U_1)^2}  b_{n}  \text{, and }\\
     b_n &\xrightarrow{U_0}{}& a_{n+1} \xrightarrow{(U_1)^2} -a_{n+1},
    \end{eqnarray*}
  for $n=1,\dots N-1$.
  So that in the subspace of $a_{n+1}$ and $b_n$ (for $n=1,\dots, N-1$) we find $\bar{U}_F=-i\sigma_y$, and similarly to the trivial case the corresponding quasi-energies $\pm \pi/(2T)$. The Majorana operators $a_1$ and $b_N$ are special:
       \begin{eqnarray*}
   a_{1} &\xrightarrow{U_0}{}& a_1 \xrightarrow{(U_1)^2} -a_1  \text{ and }\\
     b_N &\xrightarrow{U_0}{}& b_{N} \xrightarrow{(U_1)^2} - b_{N}.
    \end{eqnarray*}
  In the subspace of $a_1$ and $b_N$ we find that $\bar{U}_F$ is equal to the negative of the identity matrix  whose two eigenvalues are $-1=e^{i \epsilon T}$, and two quasi-energies $\epsilon=\pi/T$. This corresponds to MPMs localized at the first and last site of the system.
\subsubsection{4. MZM and MPM}
The phase with MZM and MPM  at the two ends of the wire is obtained with the application of $U_0^2 U_1$ (corresponding to $\lambda_0=1, \lambda_1=1/2$.) then:
       \begin{eqnarray*}
   a_{n} &\xrightarrow{(U_0)^2}& -a_{n} \xrightarrow{U_1} b_{n}  \text{ and }\\
     b_n &\xrightarrow{(U_0)^2}{}& -b_{n} \xrightarrow{U_1} -a_{n},
    \end{eqnarray*}
  for $n=2,\dots N-2$.
  In the subspace of $a_{n}$ and $b_n$ we find $\bar{U}_F=i\sigma_y$ with quasi-energies $\pm \pi/(2T)$. The Majorana operators $a_1,b_1$ and $a_N, b_N$ are special:
       \begin{eqnarray*}
   a_{1} &\xrightarrow{(U_0)^2}& a_1 \xrightarrow{U_1} -b_1,  \\
   b_{1} &\xrightarrow{(U_0)^2}& -b_1 \xrightarrow{U_1} -a_1,  \\
   a_N &\xrightarrow{(U_0)^2}& -a_{N} \xrightarrow{U_1} b_{N}, \text{ and}\\
   b_N &\xrightarrow{(U_0)^2}& b_{N} \xrightarrow{U_1} a_{N}.
    \end{eqnarray*}
  In the subspace of $a_1$ and $b_1$ we find that $\bar{U}_F=-\sigma_x$
  with eigenvalues $\mp 1=e^{i \epsilon T}$, and two quasi-energies $\epsilon=\pi/T$ and $\epsilon=0$. The corresponding eigen-oprators are $(a_1 + b_1)/\sqrt{2}$ and $(a_1-b_1)/\sqrt{2}$, respectively. Similarly, in the subspace of $a_N$ and $b_N$ we find that $\bar{U}_F=\sigma_x$
  with eigenvalues $\pm 1=e^{i \epsilon T}$, and two quasi-energies $\epsilon=0$ and $\epsilon=\pi/T$, and the corresponding eigen-oprators are $(a_1 + b_1)/\sqrt{2}$ and $(a_1-b_1)/\sqrt{2}$, respectively. We therefore find Majorana zero and $\pi$ modes as symmetric and anti-symmetric superpositions of the elementary Majorana operators at the first and last sites of the system.

 \subsection{Electrostatic driving}
The model described in the main manuscript assumes that all parameters of the Hamiltonian can be controlled in a time-dependent fashion. However, in more realistic situations, one would like to have to control fewer parameters. A particularly attractive scenario is to leave the pairing and the hopping time independent and vary only the on-site potential $\mu$ on each site, which in many potential realizations of $p$-wave superconductors is easily done. For example, in solid-state realizations, one can imagine driving the gates controlling the electrostatic environment. A more direct realization of the Kitaev chain can be implemented by a chain of superconducting quantum dots~\cite{Sau12,Fulga13,Su17}, where the potential can be tuned locally for each dot. As we show below, from a theoretical point of view tuning only the chemical potential is equally viable as the model described in the main manuscript, except that such a model does not exhibit the "sweet spot" parameters with zero correlation length for the MZMs and \MPMs{}.

In this section we study a Floquet model in which the Kitaev Hamiltonian is applied over a period $T$ where the chemical potential $\mu$ is varied from a value of $\mu_1$ in one part of the cycle to a value $\mu_2$ in the remaining part.  The Floquet operator reads:
 \begin{eqnarray}\label{eq:stroboscopic}
 U &=& e^{-iH_1 T_1} e^{-iH_2 T_2} \\
 H_j &=& \sum_i\Big[-\mu_j c_i^\dagger c_i -\frac{w}{2} \left(c_i^\dagger c_{i+1} +\text{h.c.}\right) \nonumber\\
 &+&\frac{\Delta}{2} \left(c_i c_{i+1}+\text{h.c.} \right)\Big], 
 \end{eqnarray} 
where the total Floquet period is $T=T_1+T_2$.
One can find the topological invariants of the above system by considering a ring with periodic boundary conditions and noting that at the time-reversal invariant momentum points $k=0,\pi$ the two parts of the Floquet operator commute since the order parameter vanishes.  At these points the quasi-energy is simply the time averaged kinetic energy shifted into the first Floquet zone.  This allows us to simplify the general formula of Ref.~\onlinecite{Jiang} and write:
\begin{eqnarray}\label{eq:invariants}
Q_0 &=& (-1)^{\lfloor {\overline{E_{\rm k}(k=0)}T} \rfloor+\lfloor {\overline{E_{\rm k}(k=\pi)}T} \rfloor} \nonumber \\
&=& (-1)^{\lfloor{(- \mu_1 +w )\lambda T+(- \mu_2 +w )(1-\lambda) T }\rfloor} \nonumber\\
&\times& (-1)^{\lfloor{(- \mu_1-w)\lambda T+(- \mu_2-w)(1-\lambda) T}\rfloor}  \\
Q_0\cdot Q_\pi &=& (-1)^{\lfloor {\overline{E_{\rm k}(k=0)}2T} \rfloor+\lfloor {\overline{E_{\rm k}(k=\pi)}2T} \rfloor} \nonumber\\
&=& (-1)^{\lfloor{(- \mu_1+w)\lambda 2T+(- \mu_2+w)(1-\lambda) 2T }\rfloor} \nonumber \\
&\times& (-1)^{\lfloor{(- \mu_1-w )\lambda 2T+( - \mu_2-w)(1-\lambda) 2T }\rfloor \label{eq:Q0Qpi}}.
\end{eqnarray}
Here, $\overline{E_{\rm k}(k)} = {1\over T}\int_0^T (\epsilon_k(t)-\mu(t))dt$ is the kinetic energy averaged over a period $T$ and we defined the function $\lfloor x \rfloor= \text{floor}(x/2\pi)$ that counts the number of times the band was folded back into the Floquet zone. It can be checked that $Q_0$ yields $-(+) 1$ when zero energy is intersected by an odd(even) number of bands of the kinetic energy $\overline{E_\text{k}(k)}$ folded back into the first Floquet zone. Therefore, $Q_0=-1$ corresponds to the topological phase with MZMs. In Eq.~\eqref{eq:Q0Qpi} we consider doubling the period which folds back the MPMs to zero energy. The quantity $Q_0 Q_{\pi}$ then counts the combined parity of pairs of MZMs and MPMs. Note that the first line in each invariant is more general then our stroboscopic model and can be applied to any time dependent Kitaev Hamiltonian.  In addition to the stroboscopic time dependence of Eq.~\eqnref{eq:stroboscopic}, we also consider time dependent systems where the chemical potential is of the form $\mu(t) = \mu_0 + \mu_c\cos(\Omega t)$.

\begin{figure}
	\includegraphics[width=\columnwidth]{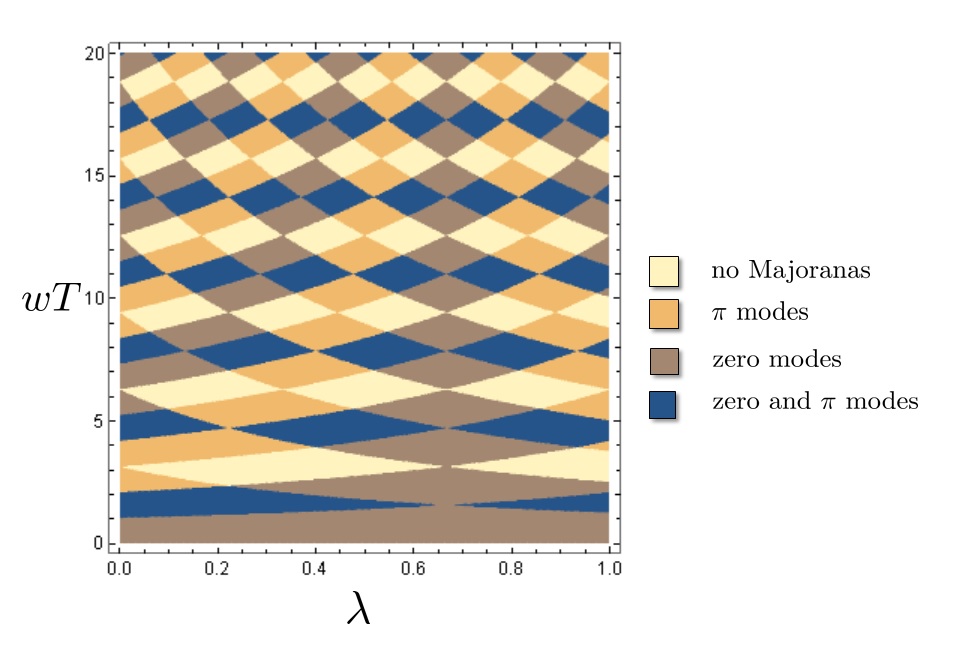}
	\caption{The phase diagram of the stroboscopic Kitaev model when only the chemical potential $\mu$ varies between two values, $\mu_1 = -0.5w$ and $\mu_2= w$. Both the total time $T$ and the relative first part of the period $\lambda= T_1/T$ are varied.}\label{fig:PD_mu_only}
\end{figure}

Fig.~\ref{fig:PD_mu_only} shows the phase diagram of the stroboscopic model when the total period $T$ is varied as well as the relative length of the first part of the period, $\lambda = T_1/T$.

\begin{figure}
	\includegraphics[width=\columnwidth]{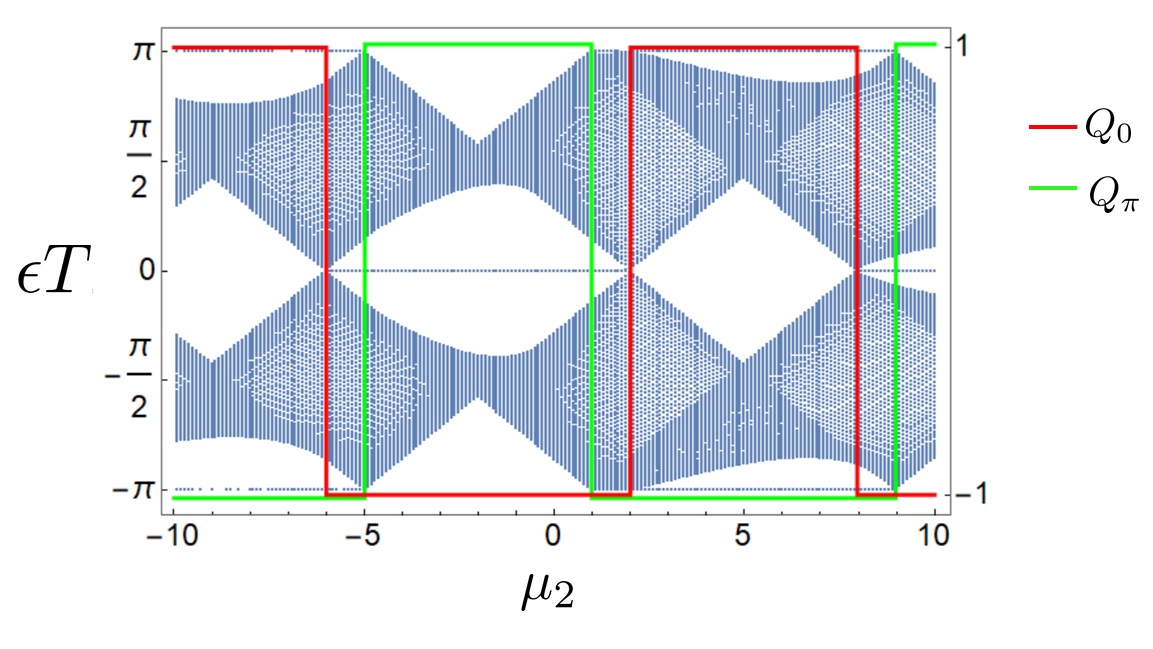}
	\caption{The quasienergy spectrum of a finite chain in the Floquet-Majorana model (left y-axis) together with the topological invariants (right y-axis). The times $T_1$ and $T_2$ are set to $0.45$, $\omega=1$, $\mu_1 = 2$ and $\mu_2$ is scanned. }\label{fig:spectrum_lambda}
\end{figure}

While Eqs.~\eqref{eq:invariants},\eqref{eq:Q0Qpi} give us the topological invariants they do not predict the size of the gap which is important for the accuracy of our procedure.  We therefore look at the stroboscopic model with an example of parameter choice where $T_1=T_2=T/2, \mu_1 = 2/T$ and varying $\mu_2$.  This gives us all three phases needed for our exchange procedure while the fourth one (a trivial phase) can be achieved by making $\mu_1=\mu_2 = 2/T$ such that the system is at the trivial equilibrium phase. Fig.~\ref{fig:spectrum_lambda} shows all quasienergies of a finite chain (of 80 sites) as a function of the changing $\mu_2$, together with the topological invariants $Q_0$ and $Q_\pi$.

\begin{figure}
	\includegraphics[width=\columnwidth]{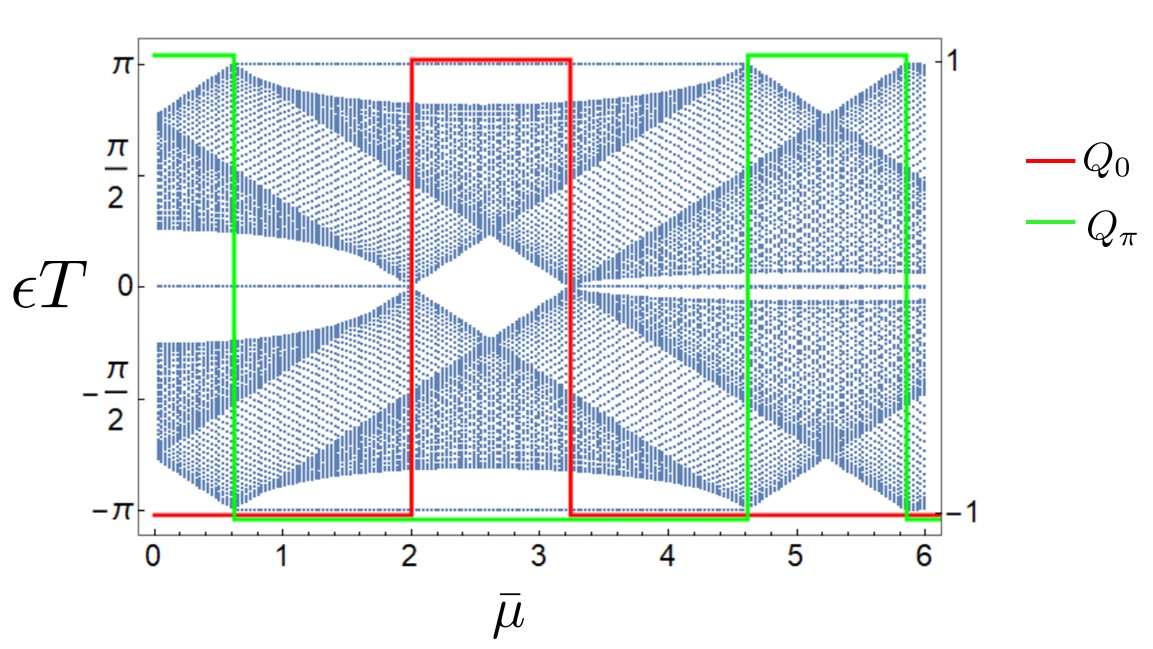}
	\caption{The quasienergy spectrum of a finite chain in the Floquet-Majorana model (left y-axis) together with the topological invariants (right y-axis).  The parameters are $T = 1.2$, $\omega=1$ and $\mu(t) = \bar{\mu} + \mu_1\cos(\Omega t)$ with $\mu_1 = 3$.}\label{fig:spectrum_cosine}
\end{figure}

Likewise we model a sinusoidal time dependent chemical potential and arrive at similar results.  The quasienergy spectrum is obtained by discretizing time, i.e. calculating the time evolution over a period as the product of evolution operators over small time slices.  The results are shown in Fig.~\ref{fig:spectrum_cosine}. 

\subsection{Numerical methods}
We now review the method by which we calculate the time evolution of the system. In any time step our Hamiltonian is bilinear in the Majorana operators $\gamma_i$ and we write its general form as 
 \begin{equation}
 H_{ij}(t) = \vec\gamma^T \bar{J} \vec \gamma 
 \end{equation}
 where $\vec \gamma$ is a column vector of Majorana operators and $\bar{J}$ is an antisymmetric imaginary matrix. (We denote matrices of c-numbers with an overbar.) The time evolution operator contains an exponent of the Hamiltonian and acts on the Majorana operators.  Let us denote by $\vec{v}_j$ the eigenvectors of $\bar{J}$ with corresponding eigenvalues $v_j$. We can express any linear combination of Majorana operators as $V= \vec V\cdot\vec\gamma = \sum_j \alpha_j \vec{v}_j\cdot \vec \gamma$.  The action of the evolution operator $$U_t = \exp\left(i t \vec\gamma^T \bar{J} \vec \gamma\right)$$ on $V$ can be written as~\cite{AssaBook}:
 \begin{eqnarray}
 U_t  V  U_t^{-1} = \left(\bar{U}_t\vec V\right)\cdot\vec \gamma \\ 
 \bar{U}_t = \exp\left(4i t\bar{J}\right),
 \end{eqnarray}  
Note that the factor of $4$ stems from the anticommutation relations of Majorana operators $\{\gamma_i,\gamma_j\}=2\delta_{ij}$.
Given the Hamiltonian in each time step, we exponentiate the matrices $\bar{J}(t)$ for each step and multiply them in the correct order to obtain the full time evolution operator.

\begin{figure}
	\includegraphics{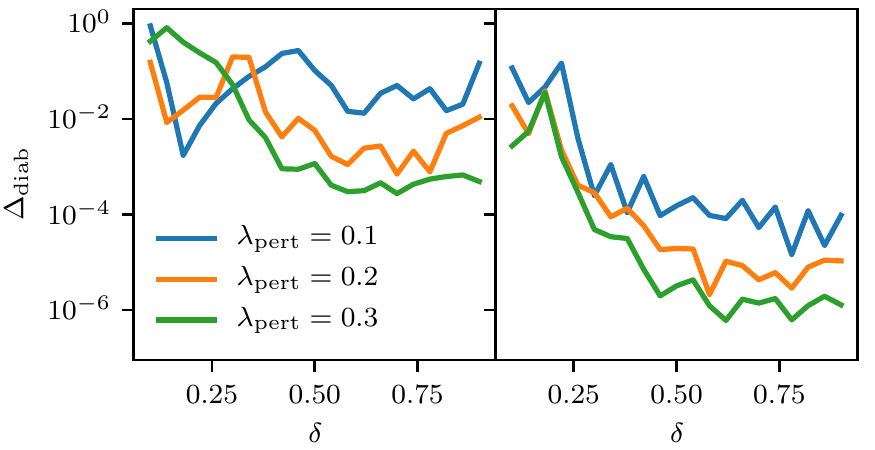}
	\caption{Diabatic errors for two different values of $N_s$ (left panel: $N_s=125$, right panel:
		$N_s = 500$) as a function of $\delta$, the deviation from the critical point, for different strengths
		of the perturbation used to split the extra pair of MZM and \MPM{}, $\lambda_{\rm pert}$.
		$\delta=1$ corresponds to the limit of vanishing correlation length. System size is $L=120$.
		It is important to note that $\delta$ controls the correlation length and the spectral
		gap, and therefore also bounds the gap induced by the perturbation. For the system size used
		here, finite-size corrections are less prevalent than diabatic errors.
		\label{fig:deltalambda} }
\end{figure}

\subsection{Parametric dependence of the diabatic errors} 

We numerically find that the parameters that control the diabatic errors -- system size $L$,
number of steps in which the modes are moved $N_s$, de-tuning from the critical point $\delta$
and perturbation strength $\lambda_{\rm pert}$ -- can exhibit very complicated interplay. Consider,
for example, the position in the phase diagram, which we control through the distance to the
critical point, $\delta$. This parameter directly or indirectly affects many physical properties of the system
and can thus have a complicated effect on the results. Its primary role is to control the spectral gap
of the unperturbed Floquet operator and the correlation length of the system. This correlation length
controls the exponent with which the hybridization between pairs of MZMs and pairs of \MPMs{} falls
off as the distance between them is increased (see also Fig.~\ref{fig:lttsb}), and thus exponentially
affects the splitting. At the same time, since it sets the gap of the undriven Floquet operator,
which also bounds the local splitting between MZMs and \MPMs{} that the perturbation can incur,
it controls diabatic corrections.

We highlight some of this complicated interplay in Fig.~\ref{fig:deltalambda}.
We observe that for small $N_s$ (left panel), the error is largely independent of $\delta$, i.e. how close
the system is to the fixed point of vanishing correlation length (which corresponds to $\delta=1$).
For larger $N_s$, the error decreases as $\delta$ is increased, i.e. the system is tuned closer to the "sweet spot".
However, in this regime we find that the dependence on system size is very weak (not shown). We
conclude from this that the finite-size errors, in particular coming from hybridization between
the MZMs and \MPMs{}, are small compared to diabatic errors. The diabatic errors are controlled by
the interplay of $N_s$ and the minimal relevant gap, which depending on the parameters can be either
the bulk gap (controlled by $\delta$) or the gap induced between MZMs and \MPMs{} in the perturbed region,
which depends on both $\delta$ and $\lambda_{\rm pert}$.

\end{document}